\documentclass[12pt,twoside]{article}
\usepackage{amsfonts,eucal}

\newtheorem{theorem}{Theorem}[section]
\newtheorem{lemma}[theorem]{Lemma}

\newtheorem{corollary}[theorem]{Corollary}

\newcommand{\C} {{\mathbb C}}
\newcommand{\N} {{\mathbb N}}
\newcommand{\R} {{\mathbb R}} 
\newcommand{\Z} {{\mathbb Z}}
\newcommand{\ca} {\mathcal{A}}
\newcommand{\cb} {\mathcal{B}}
\newcommand{\cc} {\mathcal{C}}

\newcommand{\cf} {\mathcal{F}}

\newcommand{\ebox} {V}
\newcommand{\eboxd} {\ebox'}
\newcommand{\p} {{\mathbf p}}
\newcommand{\x} {{\mathbf x}}
\newcommand{\adj} {*}
\newcommand{\ha} {a}
\newcommand{\had} {a^{\adj}}
\newcommand{\hab} {\mathbf{a}}

\newcommand{\binary} {\varepsilon}
\newcommand{\tr} {\mathrm{Tr}}
\newcommand{\z} {\zeta}
\newcommand{\tz} {\widetilde{\zeta}}
\newcommand{\avg}[1] {\left\langle {#1} \right\rangle}

\newcommand{\bra}[1] {\langle {#1} |}
\newcommand{\ket}[1] {| {#1} \rangle}

\newcommand{\skippar} {\par\medskip}
\newcommand{\qed} {\hfill {\small Q.E.D.} \par\medskip}
\newcommand{\prob} {{\mathbb P}}
\newcommand{\ds} {\displaystyle}
\newcommand{\fn} {function}

\newcommand{\pf} {partition function}
\newcommand{\gc} {grand canonical}
\newcommand{\op} {operator}
\newcommand{\pr} {probability}
\newcommand{\an} {analytic}
\newcommand{\proof} {\textsc{Proof.} }

\newcommand{\article}[3] {\textsc{{#1}}, {\itshape {#2}}, {{#3}}.}
\newcommand{\book}[3] {\textsc{{#1}}, {\itshape {#2}}, {{#3}}.}
\newcommand{\vol} {\textbf}
\newcommand{\sect}[1] {\section{{#1}} \setcounter{equation}{0}}

\renewcommand{\det} {\mathrm{det}}

\renewcommand{\r} {{\mathbf r}}
\renewcommand{\L} {\Lambda}
\renewcommand{\l} {\lambda}

\renewcommand{\hat} {\widehat}
\renewcommand{\Re} {\mathrm{Re}\,}
\renewcommand{\Im} {\mathrm{Im}\,}

\begin{document}

\title{{\scshape
	Large deviations for ideal quantum systems
}}
\author{
	Joel L. Lebowitz$^{1,2,}$\thanks{E-mail: {\ttfamily
	lebowitz@math.rutgers.edu}} \and
	Marco Lenci$^{1,}$\thanks{E-mail: {\ttfamily 
	lenci@math.rutgers.edu}} \and
	Herbert Spohn$^{3,}$\thanks{E-mail: {\ttfamily 
	spohn@mathematik.tu-muenchen.de}} \\
	\\
	$^{1}$Department of Mathematics and $^{2}$Department of Physics \\
	Rutgers University \\
	Piscataway, NJ \ 08854, U.S.A. \\
	\\
	$^{3}$Zentrum Mathematik and Physik Department \\
	Technische Universit\"at \\
	80290 M\"unchen, Germany \\
	\\
	{\scshape Dedicated to Jim McGuire} \\
	{\scshape on the occasion of his 65th birthday}
}
\date{June 1999}

\maketitle

\begin{abstract}
	We consider a general $d$-dimensional quantum system of
	non-inter\-acting particles, with suitable statistics, in a
	very large (formally infinite) container. We prove that, in
	equilibrium, the fluctuations in the density of particles in a
	subdomain $\L$ of the container are described by a large
	deviation function related to the pressure of the system. That
	is, untypical densities occur with a \pr\ exponentially small
	in the volume of $\L$, with the coefficient in the exponent
	given by the appropriate thermodynamic potential. Furthermore,
	small fluctuations satisfy the central limit theorem.
\end{abstract}

\sect{Introduction} \label{sec-intro}

Statistical mechanics is the bridge between the microscopic world of
atoms and the macroscopic world of bulk matter. In particular it
provides a prescription for obtaining macroscopic properties of
systems in thermal equilibrium from a knowledge of the microscopic
Hamiltonian. This prescription becomes mathematically precise and
elegant in the limit in which the size of the system becomes very
large on the microscopic scale (but not large enough for gravitational
interactions between the particles to be relevant). Formally this
corresponds to considering neutral or charged particles with effective
translation invariant interactions inside a container and taking the
infinite-volume or thermodynamic limit (TL).  This is the limit in
which the volume $|V|$ of the container $V$ grows to infinity along
some specified regular sequences of domains, say cubes or balls, while
the particle and energy density approach some finite limiting value
\cite{r,f,g,t}. This limit provides a precise way for eliminating
``finite size'' effects.

It is then an important result (a theorem, under suitable assumptions)
of rigorous statistical mechanics that the bulk properties of a
physical system, computed from the thermodynamic potentials via any of
the commonly used Gibbs ensembles (microcanonical, canonical, \gc,
etc.), have well defined ``equivalent'' TL's \cite{r,f,g}. These free
energy densities are furthermore proven to be the same for a suitable
class of ``boundary conditions'' (b.c.), describing the interaction of
the system with the walls and the ``outside'' of its container. When
this independence of b.c.  is ``strong enough'', the bulk free
energies also yield information about normal fluctuations, as well as
large deviations, in particle number and energy, inside large regions
of macroscopic systems. The theory of such fluctuations is at the
present time well developed for classical systems \cite{r,g,vfs}, but
almost nonexistent for quantum ones. It is the purpose of this note to
make a beginning towards such a quantum theory. This is clearly
desirable since the real world is quantum mechanical, with the
classical description being an essentially uncontrolled approximation,
albeit a very good one in many circumstances.

\subsection{Classical systems}

We begin by considering a classical system of $N$ particles of mass
$m$ in a domain, say a cubical box $V\subset \R^{d}$, interacting with
each other through a sufficiently rapidly decaying pair potential
$\phi(r)$, e.g., a Lennard-Jones potential. The Hamiltonian of the
system is then given by
\begin{equation}
	H(N,V;b) = \frac{1}{2m} \sum_{i=1}^{N} \p_{i}^{2} + \frac12
	\sum_{1\le i\ne j\le N} \phi(r_{ij}) + \sum_{i=1}^{N} 
	u_{b}(\r_{i}),
	\label{joel1.1}
\end{equation}
where $\p_{i} \in \R^{d}$, $\r_{i} \in V$, $r_{ij} = |\r_{i} -
\r_{j}|$, and $u_{b}(\r_{i})$ represents the interaction of the $i$-th
particle with the world outside of the boundary of $V$. This boundary
interaction (indicated and in the sequel by $b$) is in addition to the
action of the implicitly assumed ``hard wall'' which keeps the
particles confined to $V$. The dynamic effect of the latter is to
reflect the normal component of the particle's momentum when it hits
the wall. However, sometimes it is convenient to replace it with
periodic boundary conditions \cite{fl}.

For a macroscopic system in equilibrium at reciprocal temperature $\beta$
and chemical potential $\mu$, the \gc\ Gibbs ensemble then gives the \pr\
density for finding exactly $N$ particles inside $V\subset \R^{d}$ at the
phase point $X_{N} = (\r_{1}, \p_{1}, \ldots, \r_{N}, \p_{N}) = ({\mathbf
R}_{N}, {\mathbf P}_{N}) \in \Gamma_{N} = V^{N} \times \R^{dN}$ as
\begin{equation}
	\nu(X_{N}\,|\, \beta,\mu,V,b) = \frac{ (N!)^{-1} h^{-Nd} \exp
	[ -\beta (H(N,V;b) -\mu N) ] } {\Xi(\beta,\mu \,|\,
	V,b)}, 
	\label{joel1.3}
\end{equation}
where $\Xi$ is the \gc\ \pf
\begin{eqnarray}
	\Xi &=& \sum_{N=0}^{\infty} (N!)^{-1} \lambda_{B}^{-dN} 
	e^{\beta\mu N} \int_{V^{N}} d\r_{1} \cdots d\r_{N} \, 
	e^{ -\beta/2 \sum \phi(r_{ij}) -\beta \sum u_{b}(\r_{i}) } = 
	\nonumber \\
	&=& \sum_{N=0}^{\infty} e^{\beta\mu N} \, Q(\beta,N \,|\, V,b),
	\label{joel1.4}
\end{eqnarray}
where $Q(\beta,N \,|\, V,b)$ is the canonical \pf.  We use $h^{dN}$,
$h$ being Planck's constant, as the unit of volume in the phase space
$\Gamma_{N}$, so $\lambda_{B} = h \sqrt{\beta/(2\pi m)}$ is the de
Broglie wave length. The finite-volume, boundary condition dependent,
\gc\ pressure is
\begin{equation}
	p(\beta,\mu \,|\, V,b) = \beta^{-1} |V|^{-1} \log \Xi(\beta,\mu
	\,|\, V,b).
	\label{joel1.5}
\end{equation}

Taking now the TL, $V \nearrow \R^{d}$, we obtain, for a suitable
class of b.c., an instrinsic (b.c.~independent) \gc\ pressure
$p(\beta,\mu)$. This is related to the Helmholtz free energy density
$a(\beta,\rho)$ obtained from the TL of the canonical ensemble, i.e.,
$\Xi$ is replaced by $Q^{-1}$ in (\ref{joel1.5}) and the limit is
taken in such a way that $N/|\ebox| \to \rho$, a specified particle
density. The relation between $p$ and $a$ is given by the usual
thermodynamic formula involving the Legendre transform
\begin{equation}
	p(\beta,\mu) = \sup_{\rho} [ \rho\mu - a(\beta,\rho) ] =
	\pi(\beta,\bar{\rho}),
	\label{joel1.6}
\end{equation}
where $\pi(\beta,\rho)$ is the TL of the canonical pressure
\begin{equation}
	\pi(\beta,\rho) = -\rho^{2} \frac{\partial(a/\rho)} {\partial
	\rho}
	\label{joel1.7}
\end{equation}
and 
\begin{equation}
	\bar{\rho}(\beta,\mu) = \frac{\partial p} {\partial \mu}
	(\beta,\mu)
	\label{joel1.8}
\end{equation}
is the average density in the \gc\ ensemble. 

At a first order phase transition $p(\beta,\mu)$ is discontinuous and
the left/right limits of the derivative on the r.h.s. of
(\ref{joel1.8}) give the density in the coexisting phases. In our
discussion we shall resctrict ourselves to values of the parameters
$\beta$ and $\mu$ where the system is in a unique phase. We can of
course also go from the \gc\ pressure to the Helmholtz free energy
density by the inverse of (\ref{joel1.6}),
\begin{equation}
	-a(\beta,\rho) = \sup_{\mu} [ p(\beta,\mu) - \rho\mu ].
	\label{joel1.9}
\end{equation}

Let $P(N_{V} \in \Delta|V| \,|\, \beta,\mu,V,b)$ be the \pr\ of
finding a particle density in $V$ which lies in the interval $\Delta =
[n_{1},n_{2}]$, i.e., between the densities $n_{1}$ and $n_{2}$. Then,
for $b$ in the right class of b.c., we have (almost by definition)
that
\begin{equation}
	\lim_{V \nearrow \R^{d}} |V|^{-1} \log P(N_{V} \in \Delta|V| \,|\,
	\beta,\mu,V,b) = -\inf_{n\in \Delta} [ a(\beta,n) -
	a(\beta,\bar{\rho}) ],
	\label{joel1.10}
\end{equation}
where $\bar{\rho}$ is given by (\ref{joel1.8}). In probabilistic
language, this means that $a(\beta,n)$ is the ``large deviation
functional'' for density fluctuations. (Note that $a(\beta,\rho)$ may
be infinite for some values of $\rho$, i.e., when $\phi(r) = \infty$,
for $r < D$, and $\rho$ is above the close-packing density of balls
with diameter $D$).

On the other hand, the fluctuations in all of $V$ are clearly ensemble
dependent (they are nonexistent in the canonical ensemble) and
therefore not so physical. More relevant are the fluctuations not in
the whole volume $V$ but in a region $\L$ inside $V$.  Of particular
interest is the case when $\L$ is very large on the microscopic scale
but still very small compared to $V$. The proper idealization of this
situation is to first take the TL of $V \nearrow \R^{d}$ and then let
$\L$ itself become very large. We are thus interested in the \pr\
$P(N_{\L} \in \Delta|\L| \,|\, \beta,\mu)$, for $\L$ a large region in
an infinite system obtained by taking the TL of $\ebox$. This \pr\
should now be an intrinsic property of a uniform single-phase
macroscopic system characterized either by a chemical potential $\mu$
or by a density $\rho$.

A little thought shows that this \pr\ corresponds to considering the
\gc\ ensemble of a system of particles on a domain $\L$ with boundary
interactions of the type
\begin{equation}
	u_{b}(\r_{i}) = \sum_{k=1}^{\infty} \phi (| \r_{i} - \x_{k} |),
	\qquad \r_{i} \in V,\ \x_{k} \in V^{c},
	\label{joel1.2}
\end{equation}
i.e., we imagine that the boundary interactions come from particles of
the same type as those inside $\L$, specified to be at positions
$\x_{1}, \x_{2}, \ldots$ outside $\L$. These positions must then be
averaged according to the infinite-volume Gibbs measure. It follows
then, from the independence of the bulk properties of the system of
the boundary conditions, that equation (\ref{joel1.10}) is still
correct, that is
\begin{equation}
	\lim_{\L \nearrow \R^{d}} |\L|^{-1} \log P(N_{\L} \in \Delta|\L|
	\,|\, \beta,\mu) = -\inf_{n\in \Delta} [ a(\beta,n) - 
	a(\beta,\rho) ].
	\label{joel1.11}
\end{equation}
This relation is indeed a theorem for classical systems, under fairly
general conditions \cite{g,vfs,o}.

\subsection{Quantum systems}

It is equation (\ref{joel1.11}) and similar formulas for fluctuations
in the energy density which we want to generalize to quantum
systems. To do this, we begin by considering the boundary conditions
imposed on the $N$-particle wave \fn s $\Psi (\r_{1},\ldots, \r_{N}
\,|\, V)$ for a quantum system in the domain $V$. Usually this is done
by requiring that whenever any $\r_{i}$ is at the boundary of $V$,
$\r_{i} \in \partial V$, then $\Psi$ is equal to $\alpha$ times its
normal derivative
\begin{equation}
	\Psi (\r_{1}, \ldots, \r_{N} \,|\, V) = \alpha \: {\mathbf n}_{i}
	\cdot \frac{\partial} {\partial \r_{i}} \Psi (\r_{1}, \ldots,
	\r_{N} \,|\, V)
	\label{joel1.12}
\end{equation}
with $\alpha=0$ corresponding to Dirichlet and $\alpha=+\infty$ to
Neumann boundary conditions. 

Denote by $b_{\alpha}$ the elastic boundary condition
(\ref{joel1.12}).  The existence of the TL of the \gc\ $p(\beta,\mu
\,|\, V,b_{0})$ has been proven for quantum systems with stable
potentials \cite{r}, and for positive potentials it is established
that the pressure does not depend on $b_{\alpha}$ \cite{ro}. But, as
far as we are aware, the dependence on $u_{b}(\r_{i})$ has not been
studied systematically, with the exception of the regime covered by
the low-density expansion of Ginibre \cite{gi,br}. This only shows
that the dependence on the boundary is not so well understood for
continuous quantum systems.

To investigate the density fluctuations in quantum systems we note
that the momentum variables did not play any role in the derivation of
(\ref{joel1.10}) and (\ref{joel1.11}) for classical systems. The only
thing relevant, when considering particle number fluctuations, is the
\pr\ density in the configuration space. This is given for a classical
system by integrating $\nu$ in (\ref{joel1.3}) over the momentum
variables, whose distribution is always a product of Gaussians
(Maxwellians). For a quantum system, where the analog of
(\ref{joel1.3}) is the density matrix $\hat{\nu}$, the configuration
\pr\ density is given by the diagonal elements of $\hat{\nu}$ in the
position representation. For the \gc\ ensemble this can be written as
\begin{equation}
	\hat{W} ({\mathbf R}_{N} \,|\, \beta,\mu,V,b_{\alpha}) = \frac
	{e^{\beta\mu N} \sum_{\gamma} \left| \Psi_{\gamma} ({\mathbf
	R}_{N} \,|\, V,b_{\alpha}) \right|^{2} e^{-\beta E_{\gamma}} }
	{\hat{\Xi}},
	\label{joel1.13}
\end{equation}
where $\Psi_{\gamma}$ and $E_{\gamma}$ are the eigenstates and
eigenvalues of $H_{N}$ with the suitable statistics and $b_{\alpha}$
b.c. \cite{r,b}.

It is clear from the derivation of the TL \cite{r,f} that, when
$\phi(r)$ is super-stable, the TL for the canonical ensemble exists
for all $\rho \in [n_{1}, n_{2}]$ with b.c. $b_{\alpha}$. Then
(\ref{joel1.10}) carries over to quantum systems. The real problem is
how to prove (\ref{joel1.11}) for these systems. $\hat{W}$ is no
longer a Gibbs measure with a pair potential as interaction and there
is no good reason to expect it to be a Gibbs measure for any other
``reasonable'' many-body potential \cite{vfs}.  (Even if the latter
were the case, this potential would almost certainly depend on the
density and temperature of the system and would therefore not carry
directly any information on (\ref{joel1.11}).) It might in fact appear
that there is no strong reason why (\ref{joel1.11}) should hold for
quantum systems. The reason for expecting it to be true is that it is
a thermodynamic type relation and such relations are in general
unaffected by the transition from the classical to the quantum
formalism. More explicitly, we see the difference between
(\ref{joel1.10}) and (\ref{joel1.11}) as involving only boundary type
quantities which should become irrelevant when $\L$ is of macroscopic
size. The proof of such a statement is however far from obvious (to
us) and we therefore devote the rest of this note to proving it in the
(technically) simplest case where there are no interactions between
the particles, i.e., the ideal gas with either Bose-Einstein or
Fermi-Dirac statistics. It turns out that even in this case the proof
requires a certain amount of work.

To finish this introduction, we note that the same reasoning which
leads to (\ref{joel1.11}) also gives the well known result that the
variance of $N_{\L}$, divided by $|\L|$, is given, for $|\L| \nearrow
\infty$, by the compressibility $\avg{ (N_{\L} - \rho|\L|)^{2} } /
|\L| \to \beta^{-1} (\partial^{2} p/\partial \mu^{2})
(\beta,\mu)$. Furthermore, a central limit theorem for the random
variable $\xi = \lim_{|\L| \nearrow \infty} (N_{\L} - \rho|\L|) /
\sqrt{\rho|\L|}$ holds. These results are also expected to remain
valid for quantum systems and are proven here in the non-interacting
case.


\sect{Main results} \label{sec-formul}

We consider a $d$-dimensional square box $\ebox =
[-\ell/2,\ell/2]^{d}$. For computational convenience we choose
periodic boundary conditions. The infinite-volume thermal state we
construct below does not depend on this particular choice of the
boundary conditions, however \cite{br}.  In $\ebox$ there is an ideal
fluid (either Fermi or Bose) in thermal equilibrium, as described by
the \gc\ ensemble.  We label the Bose fluid, shorthand BE, with the
index $+$ and the Fermi fluid, shorthand FD, with the index $-$ and
introduce the Fock space
\begin{equation}
	\cf_{\pm}^{\ebox} = \C \oplus \bigoplus_{n=1}^{\infty}
	L_{\pm}^{2}(\ebox^{n})\, ,
	\label{fock}
\end{equation}
where $L_{\pm}^{2}(\ebox^{n})$ is the $n$-particle space of all
symmetric, resp. antisymmetric, square-integrable \fn s on
$\ebox^{n}$. Of course, for $n=1$, $L_{\pm}^{2}(\ebox) =
L^{2}(\ebox)$. In the sequel, in order to keep the notation light, we
will often drop sub- or superscripts whenever there is no ambiguity.

Particles do not interact. Therefore the many-particle Hamiltonian in
the box $\ebox$ can be written conveniently in the form
\begin{equation}
	H_{\ebox} = \bigoplus_{n=0}^{\infty} \sum_{i=1}^{n}
	\underbrace{1 \otimes \cdots \otimes h_{\ebox} \otimes \cdots
	\otimes 1}_{i^{th}\ \mathrm{position\ out\ of}\ n},
	\label{free-ham-1}
\end{equation}
where $h_{\ebox}$, the one-particle Hamiltonian on $L^{2}(\ebox)$, is
defined through the one-particle energy $\epsilon(k)$ in momentum
space. This means that, if $\ket{k}$ denotes the momentum eigenvector
(represented in $L^{2}(\ebox)$ as $\psi_{\ebox}^{(k)} (x) = e^{i k
\cdot x}$), then $h_{\ebox} \ket{k} = \epsilon(k) \ket{k}$ with $k \in
\eboxd = (2\pi \Z/ \ell)^{d}$, the dual of $\ebox$.

We assume $\epsilon(k)$ to be continuous, $\epsilon(0)=0$ as a
normalization, and $\epsilon(k) > 0$ for $k \ne 0$. Also $\epsilon(k)
\approx |k|^{\gamma}$ for small $k$ and $\epsilon(k) \ge |k|^{\alpha}$
for large $k$, with $\alpha,\gamma > 0$. Furthermore we require
\begin{equation}
	\int d^{d}x \, \left| \int d^{d}k \, e^{ik \cdot x} \frac1
	{e^{\beta \epsilon(k) -\beta\mu} - \binary} \right| <
	\infty   
	\label{one-p-energy}
\end{equation}
for $\binary = \pm 1$, $\beta > 0$, and suitable $\mu$.

The standard example of a non-relativistic, resp. relativistic,
kinetic energy for a particle of mass $m$ is $\epsilon(k) =
k^{2}/(2m)$, resp.\ $\epsilon(k) = \sqrt{m^{2}c^{4} + k^{2}c^{2}} -
mc^{2}$ (having set Planck's constant $\hbar=1$). Both \fn s satisfy
the above conditions. The relativistic case includes $m=0$, although
this is not immediately obvious, cf.\ Appendix \ref{subs-a0} for 
details.

We observe that $H_{V}$ may be rewritten as a quadratic form in the
creation and annihilation \op s on the Fock space $\cf$.  Let
$\had_{k}$ be the \op\ that creates a particle in the state $\ket{k}$
and $\ha_{k}$ the corresponding annihilator. Then
\begin{equation}
	H_{\ebox} = \sum_{k} \epsilon(k) \had_{k} \ha_{k} = \sum_{j,k} 
	\bra{j} h_{\ebox} \ket{k} \had_{j} \ha_{k} = 
	\bra{\hab} h_{\ebox} \ket{\hab}\,.
	\label{free-ham-2}
\end{equation}

We fix $\beta > 0$ and $\mu\in\R$ for FD, resp., $\mu<0$ for BE.  The
\gc\ state in the volume $\ebox$ is defined by
\begin{equation}
	\avg{A}_{\pm,\mu}^{\ebox} = \frac{\tr_{\cf_{\pm}^{\ebox}} \left(A
	\ e^{-\beta H_{\ebox} + \beta\mu N} \right)}
	{\Xi_{\pm}^{\ebox}(\mu)}
	\label{avg-ebox}
\end{equation}
for every bounded operator $A$ on $\cf_{\pm}^{\ebox}$. $N = N_{\ebox}$
is the \op\ for the number of particles in the box $V$,
$N|_{L_{\pm}^{2}(\ebox^{n})} = n 1_{L_{\pm}^{2}(\ebox^{n})}$, and
$\Xi_{\pm}^{\ebox}(\mu) = \tr_{\cf_{\pm}^{\ebox}} (e^{-\beta H_{\ebox}
+ \beta\mu N})$ denotes the partition function.  As is well known
(see, for example, \cite{b}) we have
\begin{eqnarray}
	\Xi_{+}^{\ebox}(\mu) &=& \prod_{k} (1 - e^{-\beta
	\epsilon(k) + \beta \mu})^{-1}\, ,  \label{pn+}  \\
	\Xi_{-}^{\ebox}(\mu) &=& \prod_{k} \left( 1 + e^{-\beta
	\epsilon(k) + \beta \mu} \right)\, .  \label{pn-}
\end{eqnarray}
The infinite-volume thermal state is defined through the limit
\begin{equation}
	\avg{\, \cdot \,} = \lim_{\ebox \nearrow \R^{d}} \avg{\, \cdot
	\,}^{\ebox} ,
	\label{avg-rd}
\end{equation}
when taking averages of local observables \cite[Sec.~2.6]{br}. 

Taking the infinite volume limit of (\ref{pn+}) and (\ref{pn-}) one 
obtains the \gc\ pressure
\begin{equation}   
	p_{\binary} (\mu) = \lim_{\ebox \nearrow \R^{d}} \frac{\log
	\Xi (\mu)} {\beta |\ebox|} = -\frac{\binary} {\beta (2\pi)^{d}} 
	\int d^{d}k \log \left(1 -\binary\, e^{-\beta \epsilon(k) + \beta   
	\mu} \right)
	\label{pressure}
\end{equation}
and the average density
\begin{equation}
	\rho_{\binary}(\mu) = \frac{dp_{\binary}} {d\mu}(\mu) =
	\frac1 {(2\pi)^{d}} \int d^{d}k \, \frac1 { e^{\beta
	\epsilon(k) -\beta\mu} - \binary}.  
	\label{density}
\end{equation}	
$p_{-}$ is real analytic on the whole axis, whereas $p_{+}$ is real
analytic only for $\mu < 0$ and has a finite limit as $\mu \to
0_{-}$. For convenience, we define $p_{+}(\mu) = \infty$ for $\mu >
0$. The slope of $p_{+}$ at $0_{-}$ is related to the Bose-Einstein
condensation. We set
\begin{equation}
	\rho_{c} = \rho(0_{-}) = \frac1 {(2\pi)^{d}} \int d^{d}k
	\, \frac1 { e^{\beta \epsilon(k)} - 1}\, .
	\label{bec-rho}
\end{equation}
By the properties of $\epsilon(k)$, $\rho_{c} = \infty$ for $d \le
\gamma$, and is finite otherwise.  $\rho_{c}$ is the maximal density
of the normal fluid and any surplus density is condensed into the $k=0$
ground state. To simplify the notation we use $\rho_{c}$ also in the
case of an ideal Fermi fluid, setting it equal to $\infty$.

The infinite system is assumed to be in a pure thermal state, obtained
through the limit (\ref{avg-rd}) at the reference chemical potential
$\mu$. In this state the average density is $\bar{\rho} = \rho(\mu) <
\rho_{c}$. We define the \emph{translated pressure} by
\begin{equation}
	g_{\binary,\mu}(\l) = g_{\binary}(\l) = p_{\binary}(\mu + \l)
	- p_{\binary}(\mu)\, .  
	\label{g-l}
\end{equation}
$g_{\binary}$ is convex up, increasing, $g(0) = 0$, and
$g_{\binary}'(0) = \bar{\rho}$. For large negative values we have
\begin{equation}
	\lim_{\l\to -\infty} g_{\binary}(\l) = -p_{\binary}(\mu) \,
	, \qquad \lim_{\l\to -\infty} g_{\binary}'(\l) = 0 ,
	\label{g1}
\end{equation}
whereas for positive values
\begin{equation}
	\lim_{\l\to \infty} g_{-}(\l) = \infty \, , \qquad \lim_{\l\to
	\infty} g_{-}'(\l) = \infty 
	\label{g3-fd}
\end{equation}
in the case of fermions and
\begin{equation}
	\lim_{\l\to -\mu} g_{+}(\l) = p_{+}(0) - p_{+}(\mu)\, , \qquad
	\lim_{\l\to -\mu} g_{+}'(\l) = \rho_{c}
	\label{g3-be}
\end{equation}
for bosons, with $g_{+}(\lambda) = \infty$ for $\lambda > -\mu$. 

We define the \emph{rate \fn} $f_{\binary}$ as the Legendre transform of
$g_{\binary}$, i.e.,
\begin{equation}
	f_{\binary,\mu}(x) = f_{\binary}(x) = \inf_{\l\in\R} \left(
	g_{\binary}(\l) - \l x \right) = g_{\binary}(\l_{o}) - \l_{o} x.
	\label{g-l0}
\end{equation}
Here $\l_{o} = \l_{o}(x)$ is the minimizer of $g(\l) - \l x$, which is
unique by convexity. For $x \le 0$ we have $\l_{o} = -\infty$.  For $0
< x < \rho_{c}$, it is determined by $g'(\l_{o}) = x$, while for $x
\ge \rho_{c}$ we have $\l_{o} = -\mu$. This shows that $f(x) =
-\infty$ on the half-line $\{ x < 0 \}$ and $f(x)$ finite
elsewhere. In particular, $f$ is convex down, strictly convex for $0 <
x < \rho_{c}$, and $f_{+}(x) = p(0) - p(\mu) + \mu x$, for $x \ge
\rho_{c}$, as a trace of the Bose-Einstein condensation.

Let us now consider a small subvolume $\Lambda$ of our (already
infinite) container $V$. The precise shape of $\Lambda$ plays no role,
only the ``surface area'' should be small compared to its volume
$|\Lambda|$. Thus, by $\L \nearrow \R^{d}$ we mean a sequence of
subdomains such that for each $\L$ there exists a subset $\L'$ of $\L$
with $|\L'|/|\L| \to 1$ and $\mathrm{dist} (\L',\R^{d} \setminus \L)
\to \infty$.

Let $N_{\L}$ be the number \op\ for the particles in $\L$. With
respect to $\avg{\, \cdot \,}$, $N_{\L}$ has some probability
distribution.  We follow the usual practice and use the same symbol
$N_{\L}$ to denote also the corresponding random variable. Its
distribution is indicated by $\prob$, averages again by $\avg{\, \cdot
\,}$.

We are now in a position to state the main result.

\begin{theorem}
	Let $\beta >0$ and $\mu<0$ for BE, resp.~$\mu\in\R$ for FD.
	Consider an interval $I=[a,b]$. Then we have, for $a <
	\rho_{c}$,
	\begin{displaymath}
		\lim_{\L \nearrow \R^{d}} \frac{1}{\beta |\L|} \log 
		\prob (\{ N_{\L} \in |\L|I \}) = \sup_{x\in I} 
		f_{\binary,\mu}(x)\, ,
	\end{displaymath}
	and, for $a \ge \rho_{c}$,
	\begin{displaymath}
		\limsup_{\L \nearrow \R^{d}} \frac{1}{\beta |\L|} \log 
		\prob (\{ N_{\L} \in |\L|I \}) \le \sup_{x\in I} 
		f_{\binary,\mu}(x)\, .
	\end{displaymath}
	\label{thm-ld}
\end{theorem}


\section{Large deviations in the density} \label{sec-large}

In this section we explain how Theorem \ref{thm-ld} follows from the
asymptotic behavior of the generating function $\avg{ e^{\beta \l
N_{\L}} }_{\binary}$.

\begin{lemma}
	There exists a $\l_{max}(\L)$ such that $\avg{ e^{\beta \l
	N_{\L}} }_{\binary} < \infty$ for all $\l < \l_{max}(\L)$ and
	$\avg{ e^{\beta \l N_{\L}} }_{\binary} = \infty$ for all $\l
	\ge \l_{max}(\L)$. For FD we have $\l_{max}(\L) = \infty$,
	whereas for BE $\l_{max}(\L) < \infty$ with $\l_{max}(\L)
	\searrow -\mu$, as $\L \nearrow \R^{d}$.  
	\label{lem-g}
\end{lemma}

\begin{theorem}
	The limit
	\begin{equation}
		\lim_{\L \nearrow \R^{d}} \frac{\log \avg{ e^{\beta
		\l N_{\L}} }_{\binary} } {\beta |\L|} = g_{\mu}(\l)\, ,
		\label{aa}
	\end{equation}
	including any finite number of derivatives, exists uniformly on 
	compacts of $\R$ for FD, resp. of $(-\infty, -\mu)$ for BE.
	\label{thm-g}
\end{theorem}

These results are proved in Sections \ref{sec-pf-lm} and
\ref{sec-pf-main}.  
\skippar

Inferring Theorem \ref{thm-ld} from our information on the generating
function is a standard argument from the theory of large deviations
\cite{e,o}.  

The probability of the event in question can be rewritten as
\begin{equation}
	Q_{\L} = \avg{ \chi_{|\L|I} (N_{\L}) } ,
	\label{t0-1}
\end{equation}
where $\chi_{A}$ is the indicator \fn\ of the set $A \subseteq \R$.
To make this event typical we introduce the modified average
\begin{equation}
	\avg{\, \cdot \,}_{\l} = \frac{1} {Z_{\l}} \avg{ \, \cdot \:
	e^{\beta\l N_{\L}}} ,
	\label{t0-2}
\end{equation}
where $\l < \l_{max}$ and the partition function $Z_{\l} =
\avg{e^{\beta\l N_{\L}}}$. With respect to this new state,
(\ref{t0-1}) can be expressed as
\begin{equation}
	Q_{\L} = Z_{\l} \, \avg{ e^{-\beta\l N_{\L}} \chi_{|\L|I}
	(N_{\L}) }_{\l} .
	\label{t0-3}
\end{equation}
An upper bound for $Q_{\L}$ comes from the exponential Chebychev
inequality,
\begin{equation}
	Q_{\L} \le \avg{ e^{\beta\l (N_{\L} - a|\L|)} } = Z_{\l} \,
	e^{-\beta\l a|\L|}
	\label{t0-7}
\end{equation}
for any $\l < \l_{max}$. 

The lower bound requires more effort and can be carried out provided
$a < \rho_{c}$.  One uses (\ref{g1})-(\ref{g3-be}) to show that there
exists a $\l_{o} < \l_{max}$ such that $g'(\l_{o}) = a$.
Differentiating (\ref{aa}) twice w.r.t.~$\l$ we obtain
\begin{eqnarray}
	& {\ds \lim_{\L \nearrow \R^{d}} \frac{ \avg{N_{\L}}_{\l_{o}}}
	{|\L|} } = g'(\l_{o}) = \rho(\mu+\l_{o}) = a\, , & 
	\label{t0-4} \\
	& {\ds \lim_{\L \nearrow \R^{d}} \frac{\beta}{|\L|} } \left[
	\avg{N_{\L}^{2} }_{\l_{o}} - \left(
	\avg{N_{\L}}_{\l_{o}} \right)^{2} \right] = {\ds
	\frac{d\rho}{d\mu} } (\mu+\l_{o})\, , &
	\label{t0-5}
\end{eqnarray}
which is finite.  This means that the event $\{N_{\L} \approx |\L| a
\}$ is typical for the new state and a law of large numbers holds.
Notice that $\l_{o}>0$, since $\rho$ is strictly increasing in
$\mu$. From (\ref{t0-3}), $\forall c\in (a,b)$,
\begin{eqnarray}
	Q_{\L} &=& Z_{\l_{o}} \avg{ e^{-\beta\l_{o} N_{\L}} 
	\chi_{|\L|[a,c]} (N_{\L}) }_{\l_{o}}  \nonumber \\
	&\ge& Z_{\l_{o}} \, e^{-\beta\l_{o} c|\L|} \, \avg{
	\chi_{|\L|[a,c]} (N_{\L}) }_{\l_{o}}   \label{t0-6} \\
	&\ge& \alpha \, Z_{\l_{o}} \, e^{-\beta\l_{o} c|\L|}  
	\nonumber 
\end{eqnarray}
for some $\alpha\in (0,1)$ and $|\L|$ large. In fact, $\avg{
\chi_{|\L|[a,c]} (N_{\L}) }_{\l_{o}} \to 1/2$ as $\L \nearrow \R^{d}$.

Therefore, when $a < \rho_{c}$, we obtain from (\ref{t0-7}),
(\ref{t0-6}) and Theorem \ref{thm-g}:
\begin{equation}
	g(\l_{o}) - \l_{o} c + o(1) \le \frac{\log Q_{\L}}
	{\beta|\L|} \le g(\l_{o}) - \l_{o} a + o(1)\, .
	\label{t0-8}
\end{equation}
Since $c \in (a,b)$ is arbitrary, we conclude that
\begin{equation}
	\lim_{\L \nearrow \R^{d}} \frac{\log Q_{\L}}{\beta|\L|} = 
	g(\l_{o}) - \l_{o} a = f(a) = \sup_{x\in [a,b]} 
	f(x)\, ,
	\label{t0-9}
\end{equation}
where $f$ is the rate \fn\ defined in (\ref{g-l0}). $\l_{o}$ is the
same as in the definition of the Legendre transform (\ref{g-l0}),
because of (\ref{t0-4}). The last equality comes from the convexity of
$f$.  

The second assertion of the theorem is now easy, since the right
inequality in (\ref{t0-8}) holds without any restriction in $a$. 
\qed

Theorem \ref{thm-ld} implies the central limit theorem for the density in
$\L$.

\begin{corollary}
	Under the assumptions of Theorem \ref{thm-ld}, the moments of
	the variable $\xi_{\L} = (N_{\L} - \avg{ N_{\L} }) /
	|\L|^{1/2}$ converge, as $\L \nearrow \R^{d}$, to those of a
	Gaussian with variance $\beta^{-1} (d\rho/d\mu) (\mu)$.
\end{corollary}
\proof The $k$-th cumulant of $\xi_{\L}$ is given by
\begin{equation}
	C_{\L}(k) = \frac1 { \beta^{k} |\L|^{k/2} } \left[
	\frac{d^{k}} {d\l^{k}} \log \avg{ e^{\beta \l N_{\L}} }
	\right]_{\l=0}\, ,
\end{equation}
$k\ge 2$. From Theorem \ref{thm-g}, $C_{\L}(2) \to \beta^{-1} g''(0) =
\beta^{-1}(d\rho/d\mu) (\mu)$, whereas, for $k > 2$, $C_{\L}(k) \to 0$. 
Also $C_{\L}(1) = 0$. These limits are the cumulants of a centered
Gaussian variable with the specified variance. 
\qed


\sect{Generating function} \label{sec-pf-lm}

We derive a determinant formula for the generating function $\avg{
e^{\beta \l N_{\L}} }_{\binary}$. With its help we prove the claims
of Lemma \ref{lem-g}. We will see in the next section that is
convenient to introduce the variables $\z = e^{\beta\lambda}$ and
$\tz = \z - 1$.

By (\ref{free-ham-2}), we have 
\begin{eqnarray}
	-\beta H_{\ebox} + \beta\mu N_{\ebox} = \bra{\hab} (-\beta
	h_{\ebox} + \beta\mu 1_{\ebox}) \ket{\hab} &=& \bra{\hab}
	A_{\ebox} \ket{\hab},  \label{t1-0}  \\
	\beta\l N_{\L} = \bra{\hab} \beta\l \chi_{\L}
	\ket{\hab} &=& \bra{\hab} B_{\L} \ket{\hab},  \label{t1-0a}
\end{eqnarray}
which implicitly define $A_{\ebox}$ and $B_{\L}$ as linear operators
on $L^{2}(\ebox)$.  We will use the following identity.

\begin{lemma}
	Let $A,B$ be self-adjoint and bounded from above. Then there
	exists a self-adjoint \op\ $C$ such that $e^{A} e^{B} e^{A} =
	e^{C}$ and 
	\begin{displaymath} 
		e^{\bra{\hab} A \ket{\hab}} e^{\bra{\hab} B \ket{\hab}} 
		e^{\bra{\hab} A \ket{\hab}} = e^{\bra{\hab} C \ket{\hab}} 
	\end{displaymath} 
	for both BE and FD.  
	\label{lemma-exp}
\end{lemma}
\proof See Appendix.
\skippar

We apply Lemma \ref{lemma-exp} with $A = A_{\ebox}/2$ and $B =
B_{\L}$, after a symmetrization of the density matrix in
(\ref{avg-ebox}). Then, using also definition (\ref{avg-rd}),
\begin{equation}
	\avg{ e^{\beta\l N_{\L}} } = \lim_{\ebox \nearrow \R^{d}}
	\frac{ \tr_{\cf^{\ebox}} ( e^{\bra{\hab} C \ket{\hab}} ) } {
	\tr_{\cf^{\ebox}} ( e^{\bra{\hab} A_{\ebox} \ket{\hab}} ) }\,.
	\label{t1-1}
\end{equation}
Evaluating the trace of a quadratic form in $\had_{i},\ha_{j}$ is a
standard calculation for both BE and FD. Let us consider first the 
case of fermions. For a self-adjoint \op\ $A$ on $L^{2}(\ebox)$
such that $e^{A}$ is trace-class, we have
\begin{equation}
	\tr_{\cf_{-}^{\ebox}} ( e^{\bra{\hab} A \ket{\hab}} ) =
	\det_{\ebox} (1_{\ebox} + e^{A}) = \det (1 + \chi_{\ebox}
	e^{A} \chi_{\ebox}), 
	\label{t1-2}
\end{equation}      
where $\det_{\ebox}$ is the determinant on $L^{2}(\ebox)$ and $\det$
the determinant on $L^{2}(\R^{d})$. Here and in the sequel we refer to
the theory of infinite determinants, as found, e.g., in
\cite[Sec.~XIII.17]{rs}.  $e^{A_{\ebox}}$ is obviously trace-class,
and so is $e^{C}$, since $e^{B_{\L}}$ is bounded.  Using the
definition of $C$, we obtain
\begin{eqnarray}
	\frac{ \det_{\ebox} (1_{\ebox} + e^{C} ) } { \det_{\ebox}
	(1_{\ebox} + e^{A_{\ebox}} ) } 
	&=&  \det_{\ebox} \left[ (1_{\ebox} + e^{A_{\ebox}})^{-1} 
	(1_{\ebox} + e^{A_{\ebox}/2} e^{B_{\L}} e^{A_{\ebox}/2}) 
	\right]  \nonumber \\
	&=& \det_{\ebox} \left[ 1_{\ebox} + (1_{\ebox} +
	e^{A_{\ebox}})^{-1} e^{A_{\ebox}/2} (e^{B_{\L}} - 1_{\ebox})
	e^{A_{\ebox}/2} \right]  \nonumber\\
	&=& \det \left[ 1 + \tz \, \chi_{\L} \,
	D_{\ebox,-} \, \chi_{\L} \right] , \label{t1-2a}
\end{eqnarray}
where $D_{\ebox,-} = (1 + e^{A_{\ebox}})^{-1} e^{A_{\ebox}}$. We used
the fact that $e^{B_{\L}} = (e^{\beta\l} - 1) \chi_{\L} + 1 = \tz
\chi_{\L} + 1$ and the cyclicity of the trace in the definition of the
determinant.  Finally, from (\ref{t1-1}) and (\ref{t1-2a}),
\begin{equation}
	\avg{ e^{\beta\l N_{\L}} }_{-} = \lim_{\ebox \nearrow \R^{d}}
	\det \left[ 1 + \tz \, \chi_{\L} \, D_{\ebox,-} \, \chi_{\L}
	\right]. 
	\label{t1-2b}
\end{equation}

One would like to take the limit on $\ebox$ inside the determinant by
replacing $D_{\ebox,-}$ with the corresponding operator on
$L^{2}(\R^{d})$ defined as
\begin{equation}
	\hat{(D_{-} \psi)} (k) = \hat{d_{-}} (k) \hat{\psi} (k), \qquad
	(D_{-} \psi) (x) = \int dy\, d_{-} (y-x) \, \psi (y),
	\label{t1-add-1}
\end{equation}
where $\ \hat{}\ $ denotes the Fourier transform and   
\begin{equation}
	\hat{d_{-}} (k) = \frac{1} {1 + e^{\beta (\epsilon(k) - \mu)}}.
	\label{t1-add-2}
\end{equation}
Notice that $\hat{d_{-}} \in L^{1} (\R^{d})$ by our assumptions on
$\epsilon(k)$ and so $d_{-} \in L^{\infty} (\R^{d})$. Moreover,
(\ref{one-p-energy}) ensures that $d_{-} \in L^{1} (\R^{d})$.

By \cite[Sec. XIII.17, Lemma 4(d)]{rs} one has to establish that
$\chi_{\L} \, D_{\ebox,-} \, \chi_{\L}$ tends to $\chi_{\L} \, D_{-}
\, \chi_{\L}$ in the trace norm.

\begin{lemma}
	Let $\hat{d}$ be a continuous integrable function on
	$\R^{d}$. We define $D$ through \emph{(\ref{t1-add-1})} 
	as a linear \op\ acting on $L^{2}(\R^{d})$. Furthermore we define
	$D_{V}$ by $D_{V} \ket{k} = \hat{d}(k) \ket{k}$ on $L^{2}(V)$
	and by $D_{V} = 0$ on the orthogonal complement $L^{2}(\R^{d} 
	\setminus V)$. Then, for $\L \subset \ebox$, $\chi_{\L}
	D_{\ebox} \chi_{\L}$ and $\chi_{\L} D \chi_{\L}$ are
	trace-class, and
	\begin{displaymath}
		\lim_{\ebox \nearrow \R^{d}} \tr |\chi_{\L} (D_{\ebox} -
		D) \chi_{\L}| = 0.
	\end{displaymath}
	\label{lemma-tr}
\end{lemma}
\proof See Appendix.
\skippar

We conclude that
\begin{equation}
	\avg{ \z^{N_{\L}} }_{-} = \det (1 + \tz \, \chi_{\L}
        \, D_{-} \, \chi_{\L} ),
	\label{t1-add-3}
\end{equation}
with $\tz = \z - 1$.
\skippar
For bosons we proceed in the same way, except that (\ref{t1-2}) is
replaced by
\begin{equation}
	\tr_{\cf_{+}^{\ebox}} ( e^{\bra{\hab} A \ket{\hab}} ) =
	\det_{\ebox} (1_{\ebox} - e^{A})^{-1}, 
	\label{t1-add-3a}
\end{equation}
requiring in addition $\| e^{A} \| < 1$. In fact, for $\| e^{A} \| \ge
1$, the l.h.s. of (\ref{t1-add-3a}) is $\infty$, whereas the
r.h.s. might be finite if 1 is not an eigenvalue of the trace-class
\op\ $e^{A}$.  In our case, by assumption $\| e^{A_{\ebox}} \| <
1$. As for $e^{C}$, the \fn\
\begin{equation}
	\l \mapsto \| e^{C} \| = \| (e^{\beta\l} - 1) e^{A_{\ebox}/2}
	\chi_{\L} e^{A_{\ebox}/2} + e^{A_{\ebox}} \|
	\label{t1-add-3b}
\end{equation}
is increasing and $\l_{max}(\L)$ is defined to be that $\l$ which
makes it equal to 1. Since the r.h.s. of (\ref{t1-add-3b}) is
increasing in $\L$ and its sup is $e^{\beta\l} \| e^{A_{\ebox}} \| =
e^{\beta (\l+\mu)}$, then one checks that $\l_{max}(\L) \searrow
-\mu$, as $\L \nearrow \R^{d}$. Therefore, following the computation
for FD, we have
\begin{equation}
	\avg{ \z^{N_{\L}} }_{+} = \lim_{\ebox \nearrow \R^{d}} \frac{ \det
	(1 - e^{C})^{-1} } { \det (1 - e^{A_{\ebox}})^{-1} } = \det (1 +
	\tz \, \chi_{\L} \, D_{+} \, \chi_{\L} )^{-1},
	\label{t1-add-4}
\end{equation}
for $\l < \l_{max}$ and $\infty$ otherwise. Here $D_{+}$, the limit of
$D_{\ebox,+} = (e^{A_{\ebox}} - 1)^{-1} e^{A_{\ebox}}$, is defined as
in (\ref{t1-add-1}) with
\begin{equation}
	\hat{d_{+}} (k)  = \frac{1} {1 - e^{\beta (\epsilon(k) -
	\mu)}}.
	\label{t1-add-5}
\end{equation}
Equation (\ref{t1-add-4}) is the analogue of (\ref{t1-add-3}) and
proves Lemma \ref{lem-g}.


\sect{Infinite volume limit} \label{sec-pf-main}

Instead of the chemical potential, in this section we use the fugacity
$z = e^{\beta\mu}$, regarding it as a complex variable. This will come
out handy for the proof of Theorem \ref{thm-g}. The variables $\z$ and
$\tz$, defined at the beginning of the previous section, will also be
extended to the complex plane. In this setup the translated pressure
(\ref{g-l}) becomes
\begin{equation}
	g_{z}(\z) = p(z\z) - p(z),
	\label{g-z}
\end{equation}
where, with a slight abuse of notation, we keep the same name for the  
pressure as a \fn\ of the fugacity. 

Expressions (\ref{pressure})-(\ref{density}) for the pressure and the
average density define two \an\ \fn s of $\mu$ in
\begin{eqnarray}
	E_{+} &=& \{ \Re \mu <0 \} \cup \{ \Re \mu \ge 0,\, \Im \mu
	\ne 2\pi j/\beta, \forall j\in\Z \},  \label{E+} \\
	E_{-} &=& \{ \Re \mu <0 \} \cup \{ \Re \mu \ge 0,\, \Im \mu
	\ne (2j+1)\pi/\beta, \forall j\in\Z \}.  \label{E-}
\end{eqnarray}
Hence $g_{\binary}(\z)$ is \an\ in
\begin{equation}
	G_{+} = \C \setminus [z^{-1},+\infty); \qquad G_{-} = \C
	\setminus (-\infty,-z^{-1}].  
	\label{G+-}
\end{equation}

We proceed to give the proof of Theorem \ref{thm-g}. Let $K \subset
G_{\binary}$ be a compact set in the complex plane. We choose $K$
such that $L = K \cap \R^{+}$ is also compact, since its image through
the \fn\ $\z \mapsto \l$ verifies the hypotheses of the theorem. Our
argument, however, is valid for any $K$. Without loss of generality,
we can assume that $1\in K$.

For $\tz$ restricted to $G_{\binary} \cap \R^{+}$, let us define
\begin{equation}
	\phi_{\binary,z}^{\L} (\z) = \frac{1}{|\L|} \log \avg{
	\z^{N_{\L}} }_{\binary,z} = -\frac{\binary}{|\L|} \tr \, 
	\log (1 + \tz \chi_{\L} D_{\binary} \chi_{\L})
	\label{t1-2d}
\end{equation}
according to (\ref{t1-add-3}) and (\ref{t1-add-4}). The proof of Theorem
\ref{thm-g} will be subdivided into three steps.
\begin{enumerate}
\item $\phi^{\L}$ can be \an ally continued to $G_{\binary}$.

\item There is a positive $r$ such that $\phi^{\L} (\z)$ converges
uniformly to $\beta g_{z}(\z)$ for $|\z-1| \le r$.

\item $|\phi^{\L}|$ is uniformly bounded on $K$. Therefore by Vitali's
lemma \cite[Sec.~5.21]{ti} $|\phi^{\L}|$ and any finite number of its
derivatives converge uniformly on $K$.
\end{enumerate}
\textsc{Step 1}. We leave the proof of the following lemma for the
Appendix.
\begin{lemma}
	The \fn\ $\phi_{\binary}^{\L}(\z)$, as defined by the trace
	in \emph{(\ref{t1-2d})}, is \an\ in $G_{\binary}$.
	\label{lemma-an}
\end{lemma}
\textsc{Step 2}. Expanding the log in (\ref{t1-2d}) one has, for 
$|\tz| < \| D_{\binary} \|^{-1}$, 
\begin{equation}
	\phi_{\binary}^{\L} (\tz+1) = -\frac{\binary}{|\L|} \tr  
	\sum_{m=1}^{\infty} \frac{(-1)^{m-1}}{m} (\tz \chi_{\L}
	D_{\binary} \chi_{\L} )^{m} .
	\label{t1-2f}
\end{equation}
We would like to interchange the summation with the trace.  To do so, we
need dominated convergence for the series:
\begin{equation}
	| \tz \chi_{\L} D_{\binary} \chi_{\L} |^{m} \le |\tz|^{m} \| 
	D_{\binary} \|^{m-1} (\chi_{\L} D_{\binary} \chi_{\L}).
	\label{t1-2g}
\end{equation}
Since $|\L|^{-1} \tr (\chi_{\L} D_{\binary} \chi_{\L}) = d_{\binary}(0)$
(see proof of Lemma \ref{lemma-tr} in the Appendix), each term of
(\ref{t1-2f}) is bounded by a term of an integrable series independent of
$\L$. Therefore, for the same $\tz$'s as above,
\begin{equation}
	\phi_{\binary}^{\L} (\tz+1) = -\binary \sum_{m=1}^{\infty} 
	\frac{(-1)^{m-1} \tz^{m}} {m} \, \frac{1}{|\L|} \, \tr (\chi_{\L} 
	D_{\binary} \chi_{\L})^{m}.
\end{equation}
Suppose that we are able to prove that
\begin{equation}
	\lim_{\L \nearrow \R^{d}} \frac{1}{|\L|} \, \tr (\chi_{\L} 
	D_{\binary} \chi_{\L})^{m} = \int dk [\hat{d}_{\binary}(k)]^{m}, 
	\label{t1-4}
\end{equation}
with a rest bounded above by $m R^{m}$ for some positive constant $R$.
Then, using (\ref{t1-add-2}) and (\ref{t1-add-5}), we would have that,
for any $r < \min \{ \| D_{\binary} \|^{-1}, R^{-1} \}$, uniformly for
$|\tz| \le r$,
\begin{eqnarray}
	\lim_{\L \nearrow \R^{d}} \phi_{\binary}^{\L} (\tz+1) &=&
	-\binary \sum_{m=1}^{\infty} \frac{(-1)^{m-1} \tz^{m}} {m}
	\int dk \left( \frac{1} {1 - \binary z^{-1} e^{\beta
	\epsilon(k)}} \right)^{m}  \nonumber \\ 
	&=& -\binary \int dk \log \left( 1 + \frac{\z - 1}{1 -
	\binary z^{-1} e^{\beta \epsilon(k)}} \right)  \\ 
	&=& -\binary \int dk \log \left( \frac{1 - \binary z\z 
	e^{-\beta \epsilon(k)}} {1 - \binary z e^{-\beta
	\epsilon(k)}} \right) = \beta \, g_{z}(\tz+1), \nonumber
\end{eqnarray}
the last equality coming from (\ref{pressure}).  This would complete
Step 2.

Let us pursue this project. One sees that
\begin{eqnarray}
	& \int dk [\hat{d}_{\binary}(k)]^{m} = \underbrace{(d
	_{\binary} * d_{\binary} * \cdots * d_{\binary})}_{m\
	\mathrm{times}} (0) & \label{t1-5} \\
	& = {\ds \frac{1}{|\L|} \int_{\L} dx_{1} \int_{\R^{d}}
	dx_{2}\,d_{\binary}(x_{1}-x_{2}) \cdots \int_{\R^{d}}
	dx_{m}\, d_{\binary}(x_{m-1}-x_{m})\,
	d_{\binary}(x_{m}-x_{1}). } & \nonumber
\end{eqnarray}
The normalized integration over $x_{1}$ is harmless since, by
translation invariance, the integrand does not depend on that
variable.  On the other hand, it is not hard to verify that
\begin{eqnarray}
	& {\ds \frac{1}{|\L|} \tr (\chi_{\L} D_{\binary}
	\chi_{\L})^{m} = \frac{1}{|\L|} \int_{\L} dx_{1} \,
	\bra{x_{1}} (\chi_{\L} D_{\binary} \chi_{\L})^{m} \ket{x_{1}}
	} & \label{t1-6} \\ 
	& = {\ds \frac{1}{|\L|} \int_{\L} dx_{1} \int_{\L} dx_{2}\,
	d_{\binary} (x_{1}-x_{2}) \cdots \int_{\L} dx_{m}\,
	d_{\binary}(x_{m-1}-x_{m})\, d_{\binary}(x_{m}-x_{1}). } &
	\nonumber
\end{eqnarray}
In view of (\ref{t1-4}), we want to compare (\ref{t1-5}) with
(\ref{t1-6}). We observe that
\begin{equation}
	\underbrace{\int_{\R^{d}} \int_{\R^{d}} \cdots \int_{\R^{d}}
	}_{m-1\ \mathrm{times}} - \underbrace{\int_{\L} \int_{\L}
	\cdots \int_{\L} }_{m-1\ \mathrm{times}} = \sum_{i=1}^{m-1}
	\, \underbrace{ \int_{\L} \cdots \int_{\L} }_{i-1\
	\mathrm{times}} \int_{\L^{c}} \underbrace{ \int_{\R^{d}}
	\cdots \int_{\R^{d}} }_{m-1-i\ \mathrm{times}}.
\end{equation}
Subtracting (\ref{t1-5}) from (\ref{t1-6}) leads then to $m-1$ terms 
of the form
\begin{equation}
	\frac{1}{|\L|} \int_{\L} dx_{1} \int_{\L^{c}} dx_{2} \,
	d_{\binary}(x_{1}-x_{2}) \cdots \int_{A_{m}} dx_{m}\,
	d_{\binary} (x_{m-1}-x_{m})\, d_{\binary}(x_{m}-x_{1}),
	\label{t1-7}
\end{equation}
where the sets $A_{3}, \ldots, A_{m}$ can be either $\R^{d}$ or
$\L$. (\ref{t1-7}) holds because, due to the cyclicity of the
integration variables, one can cyclically permute the order of
integration without touching the integrand.  We overestimate by
switching to absolute values and integrating $x_{3}, \ldots, x_{m}$
over $\R^{d}$,
\begin{equation}
	\frac{1}{|\L|} \int_{\L} dx_{1}\, (\chi_{\L^{c} - x_{1}}
	|d_{\binary}|) * |d_{\binary}| * \cdots *
	|d_{\binary}|)(x_{1}) = \frac{1}{|\L|} \int_{\L} dx_{1}
	u_{\L}(x_{1}),
	\label{t1-7a}
\end{equation}
which defines $u_{\L}(x_{1})$.  To estimate this \fn, we use
recursively the relation $\|f * g\|_{\infty} \le \|f\|_{\infty}
\|g\|_{1}$ and obtain
\begin{equation}
	u_{\L}(x_{1}) \le \|d_{\binary}\|_{1}^{m-1} \sup_{\L^{c} - x_{1}} 
	|d_{\binary}|.
	\label{t1-8}
\end{equation}
Recalling now the definition of $\L'$ given before the statement 
of Theorem \ref{thm-ld}, one sees that, if $x_{1}\in\L'$ and 
$y\in\L^{c} - x_{1}$, then $|y| \to \infty$ as $\L \nearrow \R^{d}$.  
Hence, from (\ref{t1-8}),
\begin{equation}
	\sup_{x_{1}\in\L'} \sup_{m\ge 1} \|d_{\binary}\|_{1}^{-m+1} 
	u_{\L}(x_{1}) \to 0.
\end{equation}
Also from (\ref{t1-8}), \emph{pointwise in $x_{1}$},
\begin{equation}
	\sup_{m\ge 1} \|d_{\binary}\|_{1}^{-m+1} u_{\L}(x_{1}) \le 
	\|d_{\binary}\|_{\infty}.
\end{equation}
When we average over $x_{1}\in\L$, the last two relations and 
the properties of $\L'$ prove that
\begin{equation}
	\lim_{\L \nearrow \R^{d}} \sup_{m\ge 1}
	\|d_{\binary}\|_{1}^{-m+1} \frac{1}{|\L|} \int_{\L} dx_{1}
	u_{\L}(x_{1}) =0.
\end{equation}
This takes care of each term as in (\ref{t1-7}), and we have $m-1$ of 
these terms.  Hence (\ref{t1-4}) holds with $R =
\|d_{\binary}\|_{1}$.  This ends Step 2.
\skippar\noindent
\textsc{Step 3}. Again we expand (\ref{t1-2d}) in powers of $\tz$, but
this time about a generic $\tz_{0} \not\in G_{\binary}-1$ (see
(\ref{G+-})).  We obtain
\begin{eqnarray}
	&& {\ds \frac{1}{|\L|} } \tr \, \log (1 + \tz \chi_{\L} 
	D_{\binary} \chi_{\L}) = {\ds \frac{1}{|\L|} } \tr \, \log 
	(1 + \tz_{0} \chi_{\L} D_{\binary} \chi_{\L})    
	\label{t1-10} \\
	&+& {\ds \frac{1}{|\L|} \tr \sum_{m=1}^{\infty}
	\frac{(-1)^{m-1}}{m} } \left( (1 + \tz_{0} \chi_{\L}
	D_{\binary} \chi_{\L} )^{-1} \chi_{\L} D_{\binary} \chi_{\L}
	\right)^{m} (\tz - \tz_{0})^{m}.  \nonumber
\end{eqnarray}
Let us estimate this series. First of all, using some spectral theory
\cite[Sec.~7.4]{w}, 
\begin{eqnarray}
	\| (1 + \tz_{0} \chi_{\L} D_{\binary} \chi_{\L})^{-1} \|
	&\le& \left[ \mathrm{dist} (1,\sigma(-\tz_{0} \chi_{\L}
	D_{\binary} \chi_{\L})) \right]^{-1}  \nonumber \\ 
	&\le& \left[ \mathrm{dist} (1,\sigma(-\tz_{0} D_{\binary}))
	\right]^{-1},
\end{eqnarray}
since we know from definitions (\ref{t1-add-1}), (\ref{t1-add-2}) and
(\ref{t1-add-5}) that
\begin{eqnarray}
	\sigma (\chi_{\L} D_{-} \chi_{\L}) &\subset& [0, \|\chi_{\L}
	D_{-} \chi_{\L} \|\, ] \subset [0, \| D_{-} \|\,] \nonumber \\
	&=& \sigma (D_{-}) = [ 0, 1/( 1 + z^{-1} )] , \label{t1-10a} \\
	\sigma (\chi_{\L} D_{+} \chi_{\L}) &\subset& [-\| \chi_{\L}
	D_{+} \chi_{\L} \|, 0] \subset [-\| D_{+} \|, 0] \nonumber \\
	&=& \sigma (D_{+}) = [ 1/( 1 - z^{-1} ) ,0 ].
	\label{t1-10b} 
\end{eqnarray}
Repeating the same reasoning as in Step 2, we use the above to
exchange the trace with the summation in (\ref{t1-10})---which is
legal for small $|\tz - \tz_{0}|$ as to be determined shortly.  This
yields a new series, whose $m$-th term is bounded above by
\begin{equation}
	\left[ \mathrm{dist} (1, \sigma(-\tz_{0} D_{\binary}))
	\right]^{-m} \| D_{\binary} \|^{m-1} d_{\binary}(0) \, |\tz -
	\tz_{0}|^{m} = a \left( b(\tz_{0}) \, |\tz - \tz_{0}|
	\right)^{m},
	\label{t1-11}
\end{equation}
where $d_{\binary}(0) = |\L|^{-1} \tr (\chi_{\L} D_{\binary}
\chi_{\L})$. Hence, in view of (\ref{t1-2d}), (\ref{t1-10}) implies 
\begin{equation}
	|\phi_{\binary}^{\L} (\tz+1)| \le |\phi_{\binary}^{\L}
	(\tz_{0}+1)| + a \frac{ b(\tz_{0}) \, |\tz - \tz_{0}| } { 1-
	b(\tz_{0}) \, |\tz - \tz_{0}| } \le | \phi_{\binary}^{\L}
	(\tz_{0}+1)| + a,
	\label{t1-12}
\end{equation}
for $|\tz - \tz_{0}| \le (2b(\tz_{0}))^{-1}$.  

The crucial fact is that $b(\tz)^{-1}$ stays away from zero when $\tz$
is away from the boundary of $G_{\binary}-1$.  This can be seen via
the following argument, exploiting (\ref{t1-11}) and
(\ref{t1-10a})-(\ref{t1-10b}).  In the FD case $\sigma(-\tz_{0}
D_{-})$ is a segment that has one endpoint at the origin and the phase
of $-\tz_{0}$ is the angle it forms with the positive semi-axis.  This
means that, as long as $\tz_{0}$ does not go anywhere near the
negative semi-axis, we are safe.  For $\tz_{0} \in (-z^{-1}\! -1,0)$
(see (\ref{G+-})), $\sigma(-\tz_{0} D_{-})$ is contained in
$\R_{o}^{+}$. However, notice from (\ref{t1-10a}) that the other
endpoint is located at $-\tz_{0} / (1+z^{-1}) < 1$.  For BE the
reasoning is analogous, except that in this case the phase of
$\tz_{0}$ is the angle between $\sigma(-\tz_{0} D_{+})$ and
$\R_{o}^{+}$. Therefore the ``safe'' span is the complement of the
positive semi-axis. Also, if $\tz_{0} \in (0,z^{-1}\! -1)$ (again see
(\ref{G+-})), the ``floating'' endpoint of $\sigma(-\tz_{0} D_{+})$ is
found at $\tz_{0} / (z^{-1}\! -1) < 1$.

With the above estimate we can use (\ref{t1-12}) recursively. If
$|\tz_{0}| \le r$, from Step 2, $| \phi_{\binary}^{\L} (\tz_{0}+1) |
\le M$, for some $M$, since $\phi_{\binary}^{\L}$ converges uniformly
there.  Then, from (\ref{t1-12}), we have that $| \phi_{\binary}^{\L}
(\tz_{1}+1) | \le M+a$, for any $\tz_{1}$ such that $|\tz_{1} -
\tz_{0}| < (2b(\tz_{0}))^{-1}$.  Proceeding, we see that
$|\phi_{\binary}^{\L} (\tz_{k}+1) | \le M+ka$, whenever $|\tz_{k} -
\tz_{k-1}| < (2b(\tz_{k-1}))^{-1}$.  In this way we will cover $K$ in
finitely many steps since it keeps at a certain distance from the
boundary of $G_{\binary}$ and the $(b(\tz_{k}))^{-1}$ are bounded
below.  This completes Step 3, i.e., $\phi_{\binary}^{\L} (\z)$ is
bounded on $K$ and Vitali's lemma can be applied.  
\qed


\bigskip\medskip

\noindent
{\large \textbf{Acknowledgements}}. We thank S.~Olla and G.~Gallavotti
for very instructive discussions. Work at Rutgers was supported by NSF
Grant DMR-9813268.

\medskip


\appendix
\sect{Appendices} 

\subsection{Relativistic massless particles} \label{subs-a0}

We prove that the energy dispersion $\epsilon(k) = c|k|$ satisfies our
assumptions. The only condition to be checked is (\ref{one-p-energy}),
that is, the Fourier transform of $k \mapsto ( e^{\beta(c|k| -\mu)} -
\binary )^{-1}$ is in $L^{1}(\R^{d})$. This is a consequence of the
following

\begin{lemma}
	Let $f: [0,+\infty) \longrightarrow \C$ be of Schwartz class.
	With the common abuse of notation, denote by $\hat{f}(|\xi|)$ 
	the Fourier transform of $f(|x|)$, for $x,\xi \in \R^{d}$.  
	Then, for some positive $C$, 
	\begin{displaymath} 
		\hat{f} (|\xi|) \le \frac{C} { |\xi|^{d+1} }. 
	\end{displaymath}
\end{lemma}
\proof For simplicity let us write $\xi = |\xi|$. The Fourier
transform of a radial \fn\ is
\begin{equation}
	\hat{f}(\xi) = \frac{ (2\pi)^{d/2} } { \xi^{d/2-1} }
	\int_{0}^{\infty} dr \, f(r) \, r^{d/2} J_{d/2-1} (r\xi)\, ,
	\label{radial-1}
\end{equation}
cf. \cite[Chap.~IV, Th.~3.3]{sw}, where $J_{\nu}$ is the standard Bessel
\fn\ of order $\nu$ \cite{wa}. One has
\begin{equation}
	J_{\nu}(x) \approx \frac{x^{\nu}} {2^{\nu} \Gamma(\nu+1)}, 
	\label{radial-2}
\end{equation}
for $x \to 0$, whereas 
\begin{equation}
	J_{\nu}(x) = \sqrt{ \frac2{\pi x} } \, \left[ \cos \left( x -
	\frac{2\nu + 1}{4} \, \pi \right) + g(x) \right], 
	\label{radial-3}
\end{equation}
with $g(x) \to 0$ for $x \to \infty$. Using the relation
\begin{equation}
	\int_{0}^{x} dt \, t^{\nu} J_{\nu-1} (t) = x^{\nu} 
	J_{\nu} (x)
	\label{radial-4}
\end{equation}
we integrate (\ref{radial-1}) by parts repeatedly, taking into account
also (\ref{radial-2}) and the hypothesis on $f$. After $n$
integrations we get, up to constants, $n$ terms of the form
\begin{equation}
	\frac1 {\xi^{d/2+n-1}} \int_{0}^{\infty} dr \, f^{(i)}(r) \,
	r^{d/2-n+i} J_{d/2+n-1} (r\xi),
	\label{radial-5}
\end{equation}
with $i = 1, \ldots, n$. For our purposes it suffices to iterate up to
$n \ge d/2+2$. In fact, if $i$ is such that $d/2-n+i > -d$, then in
(\ref{radial-5}) we can estimate the Bessel \fn\ by a constant. The
integral converges by the rapid decay of $f^{(i)}$ and the whole term
is of the order $\xi^{-d-1}$ or better.  For smaller values of $i$,
the estimate uses (\ref{radial-2}), for $x \in [0,a]$, and
(\ref{radial-3}) otherwise. Since $|f^{(i)}| \le c$, (\ref{radial-5})
is bounded by
\begin{eqnarray}
	&& \frac{A} {\xi^{d/2+n-1}} \int_{0}^{a/\xi} dr \, r^{d/2-n+i}
	(r\xi)^{d/2+n-1} + \nonumber \\
	&+& \frac{B} {\xi^{d/2+n-1}} \int_{a/\xi}^{\infty}
	dr \, r^{d/2-n+i} (r\xi)^{-1/2} \approx \frac1 {\xi^{d+i}},
\end{eqnarray}
the second integral being convergent because of the choice of $i$.
\qed

\subsection{Proof of Lemma 4.1} \label{subs-a1}

As before, we set $\binary = \pm 1$, according to either bosons or
fermions.  A general $f \in L^{2}(\ebox)$ can be expanded in the
Fourier basis as $f = \sum_{k} f_{k} \ket{k}$. The corresponding
creation \op\ is then defined by
\begin{equation}
	a(f)^{\adj} = \sum_{k\in\eboxd} f_{k} \, \had_{k}.
	\label{ad-f}
\end{equation}
For the sake of simplicity, we denote $\ca = \bra{\hab} A \ket{\hab} =
\sum_{ij} A_{ij} \had_{i} \ha_{j}$ (same for $\cb$). Recalling the
canonical (anti)commutation relations,
\begin{eqnarray}
	[ \ha_{i}, \had_{j} ]_{-\binary} &=& \ha_{i} \had_{j} -
	\binary\, \had_{j} \ha_{i} = \delta_{ij}; \nonumber \\ \relax
	[ \ha_{i}, \ha_{j} ]_{-\binary} &=& \ha_{i} \ha_{j} - 
	\binary\, \ha_{j} \ha_{i} = 0, \label{ccr-car}
\end{eqnarray}
one calculates that
\begin{equation}
	\left[ \ca, a(f)^{\adj} \right] = a(Af)^{\adj},
\end{equation}
and in exponential form
\begin{equation}
	e^{t\ca} a(f)^{\adj} e^{-t\ca} = a(e^{tA} f)^{\adj}.
	\label{lemma-exp-2}
\end{equation}

Now, let $\ket{0}$ be the ground state of $\cf^{\ebox}$. For $n\in\N$, and
$f_{1},f_{2}, \ldots, f_{n} \in L^{2}(\ebox)$, the finite linear
combinations of the states
\begin{equation}
	\ket{f_{1},f_{2}, \ldots, f_{n} } = a(f_{1})^{\adj}
	a(f_{2})^{\adj} \cdots a(f_{n})^{\adj} \ket{0}
	\label{lemma-exp-3}
\end{equation}
are dense in $\cf$, which is another way of stating that $\ket{0}$ is
cyclic w.r.t. the algebra generated by the creation \op s. Therefore, we
need only test our assertion on vectors of the type (\ref{lemma-exp-3}).
Using (\ref{lemma-exp-2}) with $t=1$, and observing that $\ca \ket{0} =
0$, we obtain
\begin{eqnarray}
	e^{\ca} \ket{f_{1}, \ldots, f_{n} } &=& e^{\ca} a(f_{1})^{\adj}
	e^{-\ca} \cdots e^{\ca} a(f_{n})^{\adj} e^{-\ca} \ket{0} 
	\nonumber \\
	&=& a(e^{A} f_{1})^{\adj} \cdots a(e^{A} f_{n})^{\adj} \ket{0} 
	\label{lemma-exp-4} \\
	&=& \ket{e^{A} f_{1}, \ldots, e^{A} f_{n} }.
	\nonumber
\end{eqnarray}
The existence of $C$ is a consequence of the spectral theorem. We call
$\cc$ the corresponding quadratic form in $\had_{i}, \ha_{j}$. Through
the repeated use of (\ref{lemma-exp-4}), one checks that applying
$e^{\ca} e^{\cb} e^{\ca}$ to the states (\ref{lemma-exp-3}) is the
same as applying $e^{\cc}$.  The semiboundedness of $A$ and $B$
ensures that the domain of their exponentials is the whole
$L^{2}(\ebox)$ and all quantities are well defined.  
\qed

\subsection{Proof of Lemma 4.2} \label{subs-a2}

For any symmetric \op\ $A$, $\chi_{\L} A \chi_{\L} \le \chi_{\L} |A|
\chi_{\L}$. Hence $|\chi_{\L} A \chi_{\L}| \le \chi_{\L} |A|
\chi_{\L}$ and $\tr |\chi_{\L} A \chi_{\L}| \le \tr_{\L} |A|$. When
$A = D_{\ebox}$, the convergence of the trace is proven by writing the
further estimate $\tr_{\L} |D_{\ebox}| \le \tr_{\ebox} |D_{\ebox}|$
and then summing an intregrable sequence of discrete eigenvalues. For
$A = D$, one uses the Dirac-delta representation of the trace to find
out that $\tr_{\L} |D| = |\L| \, (2\pi)^{-d} \int |\hat{d}|$. The
first assertion of the lemma has been proven.

As for the second part, let us write
\begin{eqnarray}
	&& \tr (|\chi_{\L} (D_{\ebox} - D) \chi_{\L}|) = \tr (U \chi_{\L}
	(D_{\ebox} - D) \chi_{\L})  \label{lemma-tr-1} \\
	&=& \tr_{\ebox} (U \chi_{\L} D_{\ebox} \chi_{\L}) - \tr (U
	\chi_{\L} D \chi_{\L}) = T_{\ell} - T, \nonumber
\end{eqnarray}
where $U$ is the partial isometry $L^{2}(\L) \longrightarrow
L^{2}(\L)$ that realizes the spectral decomposition as in
\cite[Th.~IV.10]{rs}.  It is convenient to use the position
representation for the bases. So, $\psi^{(k)}(x) = e^{ik\cdot x}$ and,
as defined in Section \ref{sec-formul}, $\psi_{\ebox}^{(k)} =
\psi^{(k)} \chi_{\ebox}$.  Let us work on $T_{\ell}$: using the
cyclicity of the trace one obtains
\begin{eqnarray} 
	T_{\ell} &=& \frac1 {\ell^{d}} \sum_{k \in \eboxd} \bra{
	\psi_{\ebox}^{(k)} } \chi_{\L} U \chi_{\L} D_{\ebox} \ket{
	\psi_{\ebox}^{(k)} }  \nonumber \\
	&=& \frac1 {\ell^{d}} \sum_{k \in \eboxd} \hat{d}(k)\, \bra{
	\psi_{\ebox}^{(k)} } \chi_{\L} U \chi_{\L} \ket{
	\psi_{\ebox}^{(k)} }  \label{lemma-tr-2} \\
	&=& \frac1 {\ell^{d}} \sum_{k \in \eboxd} \hat{d}(k)\, \bra{
	\psi^{(k)} } \chi_{\L} U \chi_{\L} \ket{ \psi^{(k)} } ,
	\nonumber
\end{eqnarray}
the last equality being due to the presence of the indicator \fn s
$\chi_{\L}$. In complete analogy with the above,
\begin{equation}
	T = \frac1 {(2\pi)^{d}} \int dk \, \hat{d}(k)\, \bra{  
	\psi^{(k)} } \chi_{\L} U \chi_{\L} \ket{ \psi^{(k)} }.
	\label{lemma-tr-3}
\end{equation}
Since $| \bra{ \psi^{(k)} } \chi_{\L} U \chi_{\L} \ket{ \psi^{(k)} } |
\le |\L|$, it is obvious that (\ref{lemma-tr-2}) tends to
(\ref{lemma-tr-3}) for $\ell\to\infty$.
\qed

\subsection{Proof of Lemma 5.1} \label{subs-a3}

With regard to (\ref{t1-add-3}) and (\ref{t1-add-4}), $\det (1 + \tz
\chi_{\L} D_{\binary} \chi_{\L} )$ is entire in $\tz$ (hence in $\z$)
by \cite[Sec.~XIII.17, Lemma 4(c)]{rs}. In order to evaluate its log
(on the suitable Riemann surface) we need to avoid the zeros. Using
\cite[Th.~XIII.106]{rs}, we want to make sure that $\sigma( -\tz
\chi_{\L} D_{\binary} \chi_{\L} )$ does not hit 1. Step 3 in Section
\ref{sec-pf-lm} (see in particular formulas
(\ref{t1-10a})-(\ref{t1-10b}) and the last paragraphs) shows that this
is never the case if $\tz \not\in (-\infty, -z^{-1}-1]$, for FD, or
$\tz \not\in [z^{-1}-1, +\infty)$, for BE.
\qed

Actually, we can say more. Consider FD, just to fix the ideas.  We see
from (\ref{t1-10a}) that the ``floating'' endpoint of $\sigma( -\tz
\chi_{\L} D_{-} \chi_{\L} )$ is strictly contained in the segment
$(0,-\tz / (1+ z^{-1}))$, which means that $\tz$ is allowed to exceed
slightly $G_{-}-1$, as given by (\ref{G+-}), without any vanishing of
(\ref{t1-add-3}). The above, and an analogous argument for BE, prove
that for each finite $\L$ the domain of \an ity of
$\phi_{\binary}^{\L}(\z)$ is indeed strictly bigger than
$G_{\binary}$.

For the bosonic case this fact is related to Lemma \ref{lem-g}. In a
few words, $\chi_{\L} D_{+} \chi_{\L}$ can be thought of as defining a
Hamiltonian $h'_{\L}$ in $L^{2} (\L)$, via the relation
\begin{equation}
	\chi_{\L} D_{+} \chi_{\L} = ( e^{-\beta h'_{\L} + \beta\mu
	1_{\L} } - 1 )^{-1} \, e^{-\beta h'_{\L} + \beta\mu 1_{\L} }.
\end{equation}
(Compare this with the definition of $D_{\ebox,+}$ given after formula
(\ref{t1-add-4})). Then all the calculations we have carried out in
Section \ref{sec-pf-main} are about the \gc\ ensemble of a system of
bosons on $\L$, with energy \op\ $h'_{\L}$ and chemical potential $\l
+ \mu$.  The ground state of $h'_{\L}$ is strictly positive, in
analogy to the Hamiltonian with Dirichlet b.c. In that case the upper
bound for the chemical potential, $\l_{max}(\L) + \mu$, is stricly
bigger than zero \cite{b,zuk,vlp}.

\subsection{The case $\L = \ebox$} \label{subs-a4}

We discuss the less physical case of the large deviations for the
density in the ``large box'' $\ebox$.

Instead of first taking the limit $\ebox \nearrow \R^{d}$ of the state
$\avg{ \,\cdot\, }_{\mu}^{\ebox}$, and then looking at the asymptotic
properties of the quantity $N_{\L} / |\L|$, for $\L \nearrow \R^{d}$,
we now consider the large deviations for the variable $N_{\ebox} /
|\ebox|$ w.r.t. $\avg{ \,\cdot\, }_{\mu}^{\ebox}$, subject to the
\emph{single} limit $\ebox \nearrow \R^{d}$. We call this the case $\L
= \ebox$.

We immediately see that Theorem \ref{thm-g} is a trivial identity in
this setup,
\begin{eqnarray}
	\lim_{\ebox \nearrow \R^{d}} \frac{\log \avg{ e^{\beta \l
	N_{\ebox}} }_{\mu}^{\ebox} } {\beta |\ebox|} &=& \lim_{\ebox
	\nearrow \R^{d}} \frac{\log \Xi^{\ebox}(\mu+\l) - \log
	\Xi^{\ebox}(\mu)} {\beta |\ebox|}  \nonumber \\
	&=& p(\mu+\l) - p(\mu) = g_{\mu}(\l).
\end{eqnarray}
Therefore the equivalent of Theorem \ref{thm-ld} follows in the same 
way as outlined in Section \ref{sec-large}.

In the case $\L = \ebox$, however, it turns out that we can do more
than just this. At least for some interesting cases, we can provide a
lower bound for the large deviation relation even in the BE condensation
regime, i.e. for $a \ge \rho_{c}$. This is how it is done.
\skippar

With an eye to Section \ref{sec-large}, we see that, due to
(\ref{g3-be}), the problem is that there is no \emph{fixed} $\l_{o}$
that verifies (\ref{t0-4}). In other words, no fixed value of the ``extra''
chemical potential $\l$ can be found such that the \emph{modified}
state
\begin{equation}
	\avg{ \,\cdot\, }_{\l} = \frac{ \avg{ \,\cdot\: e^{\beta\l
	N_{\ebox} } }_{\mu}^{\ebox} } { \avg{ e^{\beta\l N_{\ebox} }
	}_{\mu}^{\ebox} } = \avg{ \,\cdot\, }_{\mu+\l}^{\ebox}
\end{equation}
has an average density $a$ (compare the above definition with
(\ref{t0-2})).  On the other hand, as it is customary in the theory of
Bose-Einstein condensation (e.g., \cite{vlp} and references therein),
one can take a variable $\l_{\ebox}$ such that
\begin{equation}
	\frac{ \avg{ N_{\ebox} }_{\l_{\ebox}} } {|\ebox|} =
	\rho^{\ebox} (\mu+\l_{\ebox}) = \frac{d}{d\mu} \frac{ \log
	\Xi^{\ebox}(\mu+\l_{\ebox}) } {\beta|\ebox|} = a.
	\label{appb-1}
\end{equation}
This formula is the analogue of (\ref{t0-4}). As regards the analogue of
(\ref{t0-5}), it is not too hard to see that
\begin{equation}
	\frac{\beta}{|\ebox|} \left[ \avg{N_{\ebox}^{2} }_{\l_{\ebox}}
	- \left( \avg{N_{\ebox}}_{\l_{\ebox}} \right)^{2} \right] =
	\frac{d \rho^{\ebox}} {d\mu} (\mu+\l_{\ebox}) \approx |\ebox|.
\end{equation}
That is, the variance of $N_{\ebox} / |\ebox|$ is of the order of a
constant, at the limit. This means that our new state makes the density
$a$ average but not typical, in the sense that no law of large numbers
holds. Thus, we cannot apply estimate (\ref{t0-6}) \emph{tout court}.

However, we do not really need the density $a$ to be typical. The
aforementioned argument goes through all the same provided that, for
any $c < a$, we find an $\alpha \in (0,1)$ such that $\avg{
\chi_{|\ebox|[a,c]} (N_{\ebox}) }_{\l_{\ebox}} \ge \alpha$, for
$\ebox$ big enough.

This suggests that we look at the asymptotic distribution of
$N_{\ebox} / |\ebox|$ in the \gc\ ensemble, when the infinite-volume
on $\ebox$ is taken under the restriction (\ref{appb-1}). This is a
known object for several choices of $V$, $\epsilon(k)$ and $d$. For
instance, when $V$ is a three-dimensional box (periodic, Neumann and
Dirichlet b.c. apply) which expands isotropically and the energy is
the non-relativistic $\epsilon(k) = k^{2}/(2m)$, it is called the
\emph{Kac distribution}. Details of the computation can be found in
\cite{zuk}, whereas accounts or generalizations of the result are
presented in many other works \cite{c,vlp,vll}. It turns out that, for
$a > \rho_{c}$,
\begin{equation}
	\lim_{\ebox \nearrow \R^{d}} \avg{ \chi_{|\ebox|J} (N_{\ebox})
	}_{\l_{\ebox}} = \frac1 {a-\rho_{c}} \int_{J \cap [\rho_{c},
	\infty)} \exp \left\{ -\frac{x-\rho_{c}} {a-\rho_{c}} \right\}
	dx.
\end{equation}
Therefore the above claim holds true under certain assumptions and, in
analogy with (\ref{t0-8}), we can write that, for $\ebox \nearrow \R^{d}$,
\begin{equation}
	g(\l_{\ebox}) - \l_{\ebox} c + o(1) \le \frac{ \log \prob( \{
	N_{\ebox} \in |\ebox| I \} ) } {\beta|\ebox|},
\end{equation}
with $\l_{\ebox} \to -\mu_{-}$ and $c$ arbitrarily close to $a$. Hence
\begin{equation}
	\liminf_{\ebox \nearrow \R^{d}} \frac{ \log \prob( \{
	N_{\ebox} \in |\ebox| I \} ) } {\beta|\ebox|} \ge
	g(-\mu) + \mu a = f(a) = \sup_{x\in [a,b]} f(x).
\end{equation}

We have thus shown that in the case $\L = \ebox$, the rate function
for the large deviations in the density exists and is equal to the
Legendre transform of the translated pressure, even in the BE
condensation regime.

\end{document}